\documentclass[final,3p,times,authoryear,review,11pt]{elsearticle}

\usepackage{amssymb}
\usepackage{lipsum}
\usepackage{hyperref}

\newcommand{\organization}[1]{#1}
\newcommand{\addressline}[1]{#1}
\newcommand{\postcode}[1]{#1}
\newcommand{\city}[1]{#1}
\newcommand{\country}[1]{#1}

\journal{}

\begin{document}

\begin{frontmatter}

\title{Regulating radiology AI medical devices that evolve in their lifecycle}

\author[aff0]{Camila Gonz\'{a}lez}
\author[aff1]{Moritz Fuchs}
\author[aff2]{Daniel Pinto dos Santos}
\author[aff4]{Philipp Matthies}
\author[aff5]{Manuel Trenz}
\author[aff5]{Maximilian Grüning}
\author[aff0]{Akshay Chaudhari}
\author[aff0]{David B. Larson}
\author[aff2]{Ahmed Othman}
\author[aff8]{Moon Kim}
\author[aff8]{Felix Nensa}
\author[aff1]{Anirban Mukhopadhyay}

\affiliation[aff0]{
  \organization={{Department of Radiology, Stanford University},
  \addressline{300 Pasteur Drive},
  \postcode{94304},
  \city{Palo Alto, CA},
  \country{USA}}
}
\affiliation[aff1]{
  \organization={{Technical University of Darmstadt},
  \addressline{Karolinenpl. 5},
  \postcode{64289},
  \city{Darmstadt},
  \country{Germany}}
}
\affiliation[aff2]{%
  \organization={{University Medical Center Mainz},
  \addressline{Langenbeckstr. 1},
  \postcode{55131},
  \city{Mainz},
  \country{Germany}}
}
\affiliation[aff4]{%
  \organization={{Smart Reporting GmbH},
  \addressline{Gewürzmühlstr. 11},
  \postcode{80538},
  \city{Munich},
  \country{Germany}}
}
\affiliation[aff5]{%
  \organization={{Georg-August University Göttingen},
  \addressline{Wilhelmsplatz 1},
  \postcode{37073},
  \city{Göttingen},
  \country{Germany}}
}
\affiliation[aff8]{%
  \organization={{University Hospital Essen},
  \addressline{Hufelandstr. 55},
  \postcode{45147},
  \city{Essen},
  \country{Germany}}
}

\begin{abstract}

Over time, the distribution of medical image data \emph{drifts} due to factors such as shifts in patient demographics, acquisition devices, and disease manifestations. While human radiologists can adjust their expertise to accommodate such variations, deep learning models cannot. In fact, such models are highly susceptible to even slight variations in image characteristics. Consequently, manufacturers must conduct regular updates to ensure that they remain safe and effective. Performing such updates in the United States and European Union required, until recently, obtaining re-approval. Given the time and financial burdens associated with these processes, updates were infrequent, and obsolete systems remained in operation for too long. During 2024, several regulatory developments promised to streamline the safe rollout of model updates: The European \emph{Artificial Intelligence Act} came into effect last August, and the Food and Drug Administration (FDA) issued final marketing submission recommendations for a Predetermined Change Control Plan (PCCP) in December. We provide an overview of these developments and outline the key building blocks necessary for successfully deploying dynamic systems. At the heart of these regulations -- and as prerequisites for manufacturers to conduct model updates without re-approval -- are clear descriptions of data collection and re-training processes, coupled with robust real-world quality monitoring mechanisms.

\end{abstract}


\end{frontmatter}

\newpage
\section{Introduction}

Imagine a use case where an AI-supported system could be helpful in the clinical routine, such as detecting pulmonary embolism in chest computer tomography (CT) \citep{colak2021rsna}. The process from design to deployment requires a complex vertical integration from collecting data to embedding the workflow in clinical practice \citep{fuchs_closing_2023}. The product -- or \emph{Software as a Medical Device (SaMD)} \citep{IMDRF_SaMD} --would typically go through the following five stages:
\begin{enumerate}
    \item \textbf{Data collection}: First, experts gather and annotate sufficient data in an appropriate format. The database includes subjects for training and testing.
    \item \textbf{Model design and training}: A portion of the data is used to select and train a predictive model, nowadays probably with Deep Learning (DL).
    \item \textbf{Evaluation:} The model is evaluated in terms of performance and robustness, ideally across both \emph{in-distribution} and \emph{out-of-distribution} (e.g., acquired at a different institution) test subject populations.
    \item \textbf{Approval}: The product receives approval from the responsible regulatory entity. In the European Union, a \emph{notified body} \citep{european2017mdr} carries out this task, and the FDA is responsible in the United States.
    \item \textbf{Deployment}: The SaMD is rolled out in clinics.
\end{enumerate}

This process seems reasonable, and the product will initially perform well as long as the evaluation was suitable for the clinical problem. What happens, however, after one year? How about five?

Deep Learning models are notorious for their frailty when faced with distribution shifts \citep{gonzalez2022distance,hendrycks2021many}. These shifts can come from changes in the image acquisition device or parameters, choice of reconstruction algorithm, subject population, and disease manifestation, among other factors \citep{midya2018influence,van2020radiomics}. Over time, the performance of models on new data deteriorates until they are no longer reliable.

How should the manufacturer proceed when the model becomes obsolete? The simple answer is to update the data to reflect the current distribution and \emph{re-train} the model. Fortunately, many vendors of SaMD continue to collect data beyond the initial deployment \citep{van2021artificial}. These new samples are often directly acquired at the institutions where the product is used, so they are stored locally --preventing privacy breaches -- and particularly valuable for increasing the performance on the actual users of the system. Many medical AI products have only been in the market for a few years \citep{van2021artificial}, so they are yet to exhibit a critical performance deterioration. But, when they do, manufacturers can \emph{in theory} leverage the more recent cases to expand the training base and perform a new round of training. After the updated model has again undergone appropriate evaluation and approval, the new version can be released.

Why only \emph{in theory}? Because, in practice, several constraints hinder this process:

\begin{enumerate}
    \item There is a considerable \textbf{backlog of SaMD waiting for approval} in the EU due to a \textbf{lack of capacity of suitable notified bodies} \citep{notified_body_capacities}. The need to re-certify all products according to the  Medical Device Regulation (MDR) is a big factor behind this delay. In July 2023, from the 35 bodies tracked by \href{https://openregulatory.com/notified-bodies/}{\emph{Open Regulatory}}, \emph{only seven} are offering audits within one year.
    \item The \textbf{re-approval process} is both \textbf{slow and extremely expensive} \citep{makower2010fda}. In particular, small start-ups that deploy their products through more established platforms have trouble bearing the cost, so they may wait longer than necessary to roll out an update.
    \item Oftentimes, \textbf{a model cannot be re-trained from scratch} as either (a) part of the data is no longer available (due, for instance, to the stipulations for data protection posed by the GDPR \citep{voigt2017eu} in the EU and HIPAA \citep{act1996health} in the USA) or (b) the training process is too computationally demanding, such as the case for foundation models \citep{moor2023foundation}.
    \item Nowhere is it defined \textbf{\emph{when} these updates should be triggered} or \emph{who} should detect when the model performance is unreliable and re-training necessary. Even if an appropriate quality assurance system is in place, by the time a performance deterioration has been identified, it is too late: The product must be taken out of the routine until an update has been conducted.
\end{enumerate}

Considering these shortcomings of the regulatory landscape in the past years, it is not surprising that AI products have often been found to perform below the initial expectations and \emph{burdened} or \emph{only minimally aided} healthcare professionals \citep{beede2020human,shin2023impact}. It is also not surprising that only 2\% of FDA-approved AI medical devices reported being updated with new data \citep{wu2024regulating}.

Fortunately, regulatory bodies around the world are moving away from this \emph{static learning} paradigm and towards a \emph{lifecycle regulatory protocol} \citep{vokinger2021continual,vokinger2021regulating}, as outlined in the \emph{\textbf{Harmonized Rules for Artificial Intelligence (AI Act)}} for European member states \citep{european2021laying} which came into effect last August and in the FDA marketing submission recommendations for a \emph{\textbf{Predetermined Change Control Plan (PCCP)}} in the United States. These new guidelines embrace the possibility of new moderate risks that emerge when adapting models to the environment as long as the risks are adequately mitigated and outweighed by the potential benefits. 

\begin{figure}[h]
\centering
\includegraphics[width=0.8\linewidth]{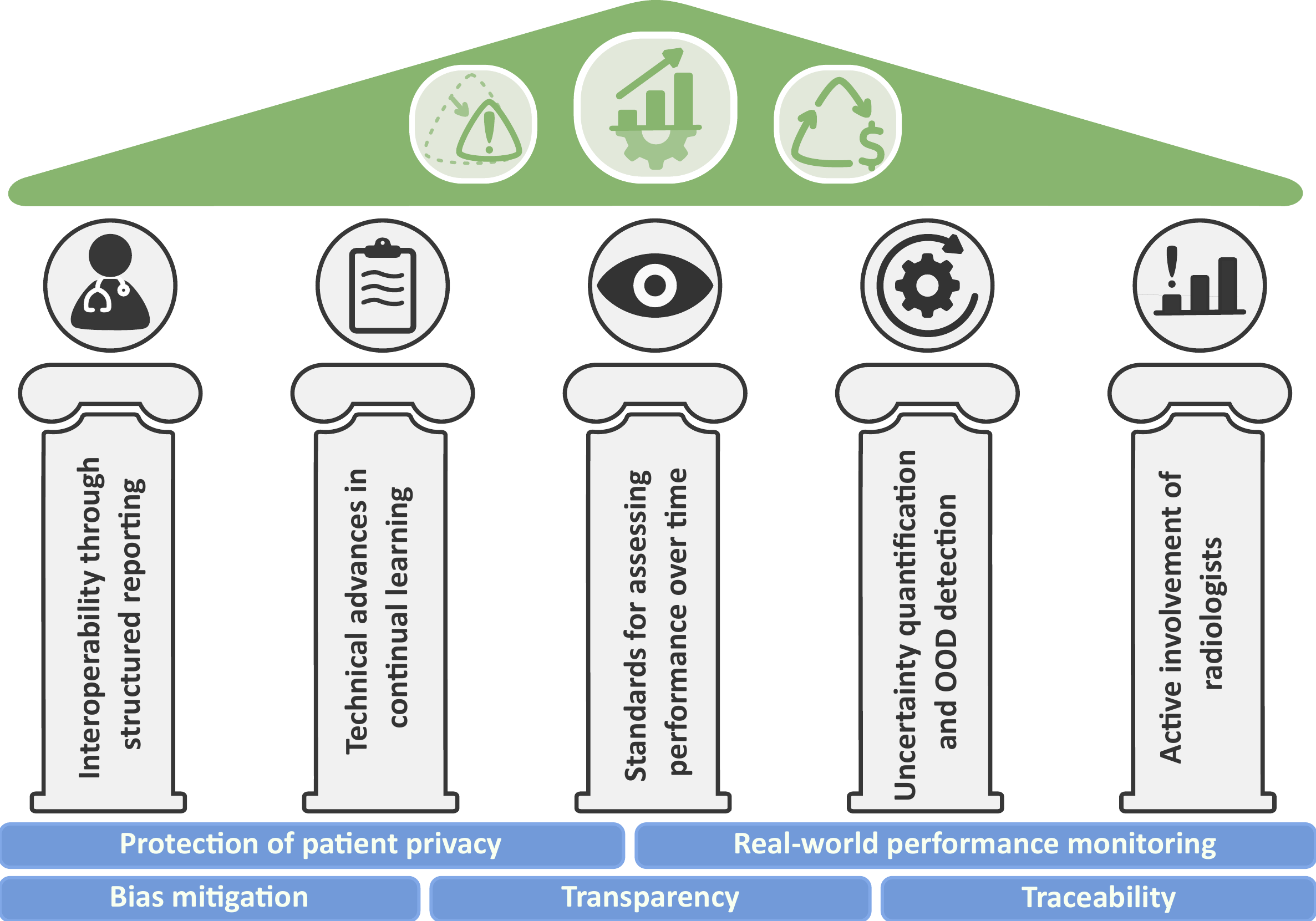}
\caption{From bottom to top: Requirements set by recent regulatory efforts in EU and USA, building blocks for effectively designing dynamic systems, and potential benefits in the form of risk reduction, performance increase, and sustainability.} \label{fig:requirements_means_objectives}
\end{figure}

Though different in certain aspects, both pose the requirements outlined in Figure \ref{fig:requirements_means_objectives} and follow the same concept, visualized in Figure \ref{fig:workflow}: when seeking approval for a SaMD, manufacturers can submit documentation on the planned changes that aim to improve the risk/benefit ratio of the product without affecting its intended use. As long as future modifications follow the specifications outlined in the protocol and fulfill the base requirements, they can be rolled out without re-approval.

In this article, we delve into the regulatory developments that have taken shape over the last few years. From this foundation, we describe the necessary elements for building systems that learn over time. We analyze how these developments can positively impact the state of AI in healthcare by reducing risks, improving performance, and allowing for sustainable business practices. We also point to some shortcomings and potential oversight in the current version of the draft guidances.

\section{Regulating dynamic systems - A brief history}

The landscape for AI regulation is rapidly evolving. In particular, several directives currently in force or under consideration in the USA and European Union address systems that are regularly updated after deployment. There are many commonalities, but also some differences, in how \emph{Software as a Medical Device (SaMD)} regulation is approached by US authorities and EU member states. One core difference is that the USA delegates decisions to sector-specific agencies, such as the \emph{Food and Drug Administration (FDA)} for medical devices. In April 2023, and after several previous efforts \citep{food2019proposed,us2021artificial}, the FDA released draft recommendations for AI-powered medical software that continues to learn based on a \emph{Predetermined Change Control Plan} (PCCP) \citep{food2023marketing}. The guidance was finalized on December 2024 \citep{FDA_PCCP_Guidance_2024}, setting new standards on how AI-based medical devices should be updated and monitored.

In contrast, manufacturers who wish to sell their products in the EU must adhere to both area-specific directives -- which currently encompass the \emph{Medical Device Regulation (MDR)} \citep{european2017mdr} that came into effect on May 26\textsuperscript{th}, 2021 -- and cross-sectional guidelines, such as the GDPR for data protection \citep{voigt2017eu} and, since August 2024, an \emph{Artificial Intelligence Act (AI Act)} \citep{european2024laying}.  If two directives are in conflict, the manufacturer must follow that which is more strictly defined or introduces the least risks. In the European case, it is the AI Act that describes how to treat continual learning within \emph{high risk} applications (such as SaMD).

\begin{figure*}[h]
\centering
\includegraphics[width=\linewidth]{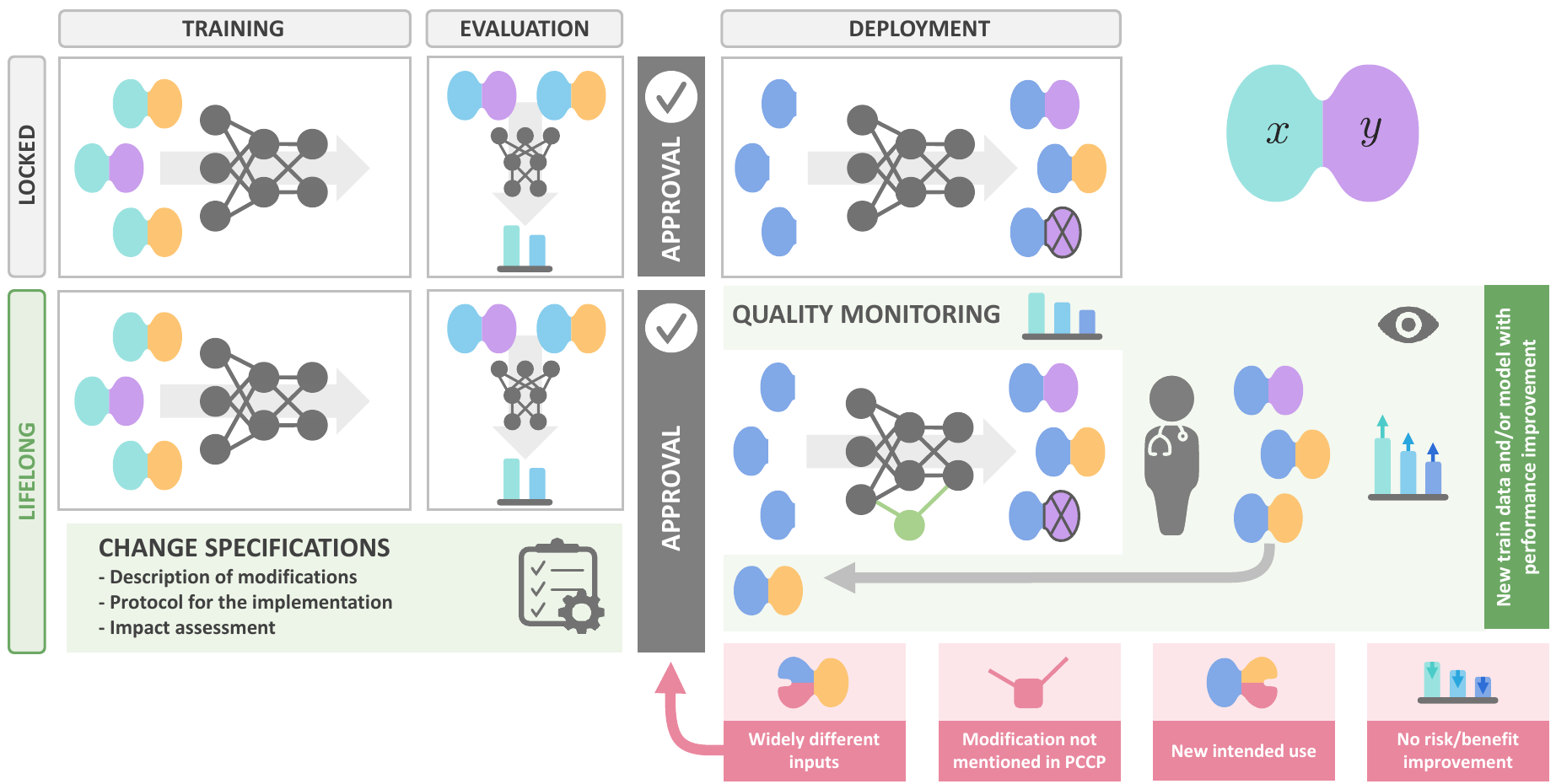}
\caption{Approval process for a deterministic \emph{locked} AI system (top) vs. a lifelong learning system (bottom). In the second case, a description of the planned modifications and protocols outlining how they will be implemented and monitored are submitted as additional documentation during initial clearance. Modifications following these specifications, such as re-training the model with additional training data, do not require re-approval.} \label{fig:workflow}
\end{figure*}

Commonalities between the two regulatory models include a new workflow -- visualized in Figure \ref{fig:workflow} -- that allows manufacturers to submit documentation on their planned modifications. Following the approval of this \emph{change control plan}, systems are able to adapt as long as they do so according to the described procedure and they remain \emph{safe and effective} for the expected conditions of use, as measured by predefined specifications. Acceptable modifications include changes to the inputs (such as compatibility with new devices), the model (for instance, after re-training or adapting the architecture), and slight adaptations in the intended use (e.g. addition of a new sub-population). However, they do \textbf{not} include changes to the intended use or that introduce new risks, particularly those affecting the risk categorization of the product. Such modifications still require a traditional pre-market review under the new regulatory framework. 

Both agencies follow the goal of mitigating high risks but embracing moderate risks as long as the potential benefits outweigh them. Without diverging from this focus, the new workflow would ensure the safety and effectiveness of the product through the mechanisms described in the following sections.

\subsection*{Real-world performance monitoring}

Both the \emph{PCCP} and \emph{AI Act} emphasize the need for a quality assurance system that continuously monitors model predictions. The system should ensure that, at all times during deployment, the predictive performance fulfills certain standards. The metrics and sub-population datasets used to measure this adherence should be described pre-approval. The risk management system should evolve dynamically and be updated regularly. Manufacturers need to identify and evaluate new risks and adopt measures to mitigate them in a timely manner.

\subsection*{Protection of patient privacy}

Medical records are highly sensitive data. In line with GDPR principles, systems should minimize the storage and transfer of patient data and adopt mechanisms to prevent data leakage. In the context of lifelong learning systems, this means that not all data used at one point for training will be available later on. Ensuring patient privacy goes beyond protecting raw data and also includes, among other safeguards, assessing how much information can be extracted from a trained model and building in mechanisms from \emph{differential privacy} \citep{kaissis2020secure}.

\subsection*{Bias mitigation}

A strong focus is placed on both directives in preventing discrimination and the generation of biased outputs. Preventing bias is challenging, as preconceptions are often internalized in the training or testing data. Opening the possibility of training with real-world data poses the risk that new biases will be incorporated into the model. To comply with this objective, the performance should be measured across distinct sub-populations, and manufacturers need to identify and correct sources of bias in both the initial data basis and new cases that may be collected later on.

\subsection*{Transparency to patients and stakeholders}

Emphasis is given to transparency towards users, regulatory bodies, and other relevant stakeholders, such as the manufacturers of all systems that communicate with the product. Both clinical users and patients should have access to information on the expected and real-life performance of the system and on the collected data. Though a difficult goal, this can be achieved through pre-registering evaluation protocols and publicly releasing the results alongside statistics on the gathered data.

Also relevant for increasing trust is providing reliable insights into a model's workings. \emph{Saliency maps} \citep{geirhos2020shortcut} are the most popular deep learning interpretability technique, though some research questions their usefulness \citep{arun2021assessing}. Other methods designed for this task include \emph{probability distribution maps} \citep{jetley2016end}, \emph{counterfactual visual explanations} \citep{goyal2019counterfactual} that illustrate what modifications to an image would alter the prediction, \emph{explanatory interactive learning} where users can provide feedback to the explanations \citep{schramowski2020making} and \emph{model-agnostic surrogate explainers} that supply patient-specific interpretations \citep{kumarakulasinghe2020evaluating}.

\subsection*{Traceability}

Key for ensuring transparency and monitoring systems performance, all modifications need to be appropriately documented in a version control system. It should be possible to reproduce any given output from the inputs and version of the product.

\section{Building an infrastructure for continuous AI}

Establishing an infrastructure that supports the requirements posed by regulatory agencies is not a trivial matter. In order to build a system that effectively and safely learns over time, manufacturers need to embrace certain components, which we elaborate on in this section.

\subsection*{Interoperability through structured reporting}

A pre-requisite for attempting any kind of model updating over time is that \emph{annotations follow a structured, predetermined format} \citep{fuchs_closing_2023}. While shifts in image acquisition, disease patterns, and patient population are unavoidable, shifts in annotation practices can -- and should -- be minimized. This means, first, abandoning narrative radiology reports for \emph{structured reporting} \citep{european2018esr}. 

Narrative reports have been found to be highly subjective, often depending on the career trajectory of the reporting radiologist, and to vary widely in terms of length and style \citep{ganeshan2018structured}. Attempting to update a model with free-text data will inevitably introduce shifts in the labeling distribution due to differences in the reporting style, even with the emergence of large language models.

Following structured reporting, a \emph{template} is first developed that determines which fields must be completed for a particular diagnosis. Radiologists then follow this template when performing a diagnosis, filling in the required fields. When all collected data follows the same format and degree of abstraction, new cases can be used to train the model without further preparation. 

Structured reporting has many advantages beyond deep learning, such as minimizing the error rate and improving communication among experts. The practice is, for instance, associated with longer survival of colorectal cancer patients \citep{sluijter2019improvement}. Of course, this correlation does not by itself indicate that improvements are due only to the use of structured reporting, as advances were also made concurrently in terms of targeted therapy to specific mutations.

Reporting in a structured fashion entails the need for a reliable infrastructure that ensures the ongoing availability of the template and smooth running of the reporting software. It also opens the question of how to proceed when changes in the template are desired or the radiologists finds the template gives no place to the information they wish to document \citep{powell2015state,ganeshan2018structured}.

\subsection*{Technical advances in continual learning}

Even when new data has been collected and labeled according to the template, adapting deep learning models is not as simple as running a few iterations with the new cases. We would ideally retrain the model jointly with all data, yet this is often not possible due to constraints on data availability or because the process is too time- or computationally expensive.

\begin{figure*}[ht]
\centering
\includegraphics[width=\linewidth]{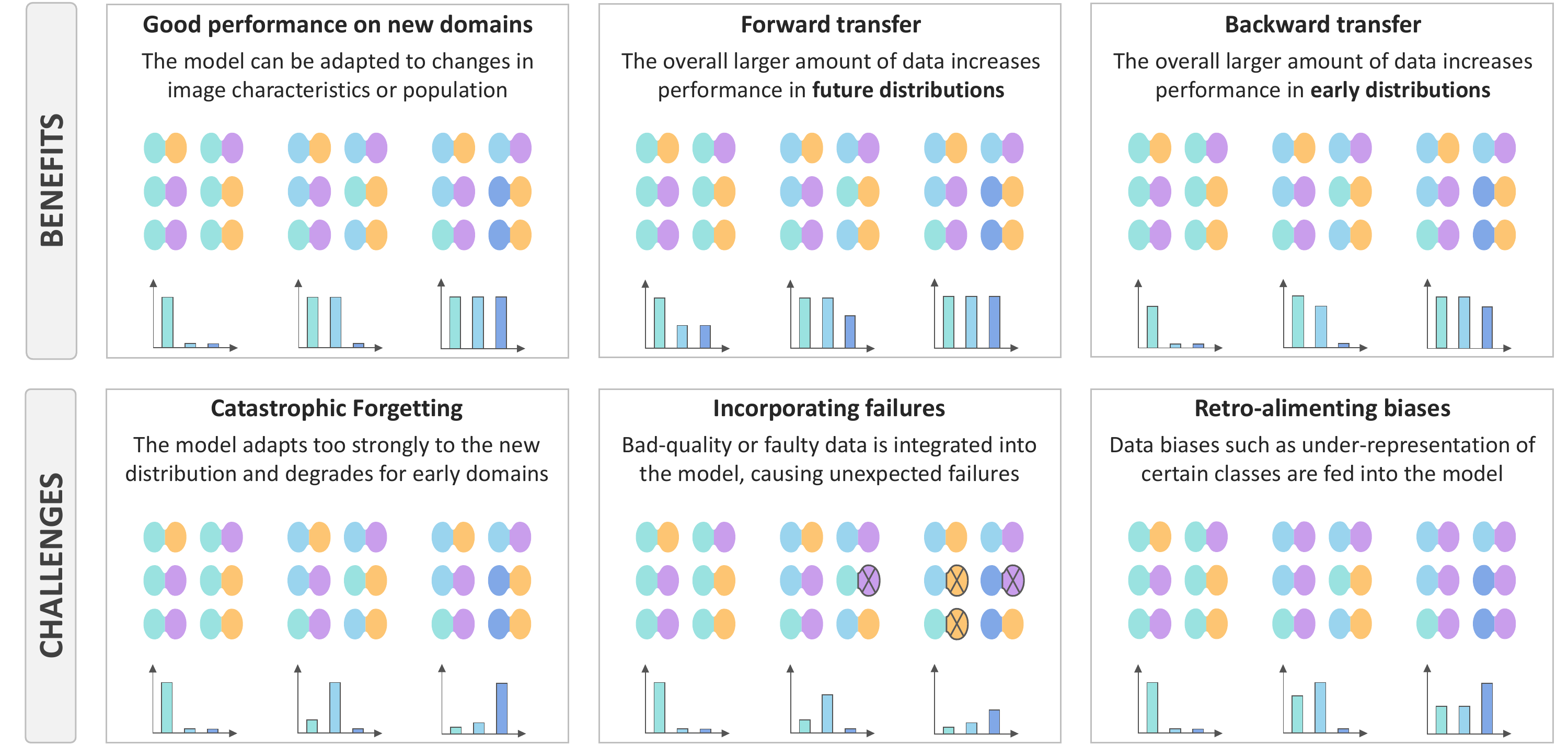}
\caption{Continuously learning systems adapt to new environments, obtaining better performance on local data distributions. By leveraging more data overall, global performance may also increase on yet-unseen or older domains. However, it is also possible that the system forgets how to cope with earlier data or that failures and/or biases are introduced into the model.} \label{fig:REG_tech_challenges_opportunities}
\end{figure*}

The goal is to leverage both the present model state -- and hereby the large data basis with which it has been trained -- and the new data. Unfortunately, if we were to simply continue training with the new cases, the model would adapt too strongly to the new data distribution and lose valuable knowledge from the previous model state. This is often referred to as \emph{catastrophic forgetting} (see Figure \ref{fig:REG_tech_challenges_opportunities}). We instead seek \emph{backward transfer}, i.e., we want our performance to improve even on data similar to that seen early on \citep{hadsell2020embracing}. We also want \emph{forward transfer}, which means greater generalization to yet-unseen acquisition practices.

Many \textit{continual learning} methods have emerged in the last few years that pursue these objectives with various success. Particularly \textit{rehearsal-based approaches}, which save a subset of previous samples and interleave them in later iterations, have been shown successful at acquiring new information whilst preserving previous knowledge \citep{perkonigg2021dynamic}. \textit{Pseudo-rehearsal} methods are a viable alternative for cases where no image data can be stored or exchanged between institutions. \textit{Ensemble-based} methods, which maintain several parametrizations of the model in tandem and select the most appropriate one during inference, have also shown promise in radiological tasks \citep{gonzalez2022task}.

\subsection*{Standards for assessing performance over time}

Though the need for a real-world performance monitoring system is highlighted by the \emph{PCCP} and \emph{AI Act}, what constitutes an increase or a deterioration in performance is not clarified. 

The following aspects should be defined by the manufacturer pre-approval:
\begin{itemize}
    \item A catalog of metrics relevant to the clinical use case that measure performance, robustness, and risks.
    \item How metrics are weighted, and how to proceed if the performance decreases according to some metrics in certain situations, but overall an improvement is shown.
    \item Strategies to collect and separate test cases and the process describing how to proceed if a new model state is better for some subject groups and/or findings and worse for others.
    \item Clear boundaries that define a substantial change in model performance and statistical tests used to corroborate a significant difference.
\end{itemize}

All these items should be defined in a clear fashion and implemented in the system.

\subsection*{Uncertainty quantification and out-of-distribution detection}

Regardless of how well a model performs, it will inevitably produce some failures. This does not signify a risk as long as the system communicates that it is uncertain about potentially faulty predictions. Such uncertainty estimates cannot -- in many cases -- be derived directly from the model outputs. Besides their susceptibility to distribution shift, another shortcoming of deep learning models is their tendency to produce overconfident predictions \citep{mehrtash2020confidence}.

A simple and reliable strategy to quantify uncertainty is with \emph{Deep Ensembles} by training several models and calculating the divergence between their predictions \citep{lakshminarayanan2017simple}. Some methodological variations decrease the computational burden by reducing the number of models or independent parameters \cite{fuchs2021practical,mehrtens2022improving}. Another strategy is through \emph{out-of-distribution detection}, by identifying images so far from the training distribution that the model can likely not produce high-quality outputs \citep{gonzalez2022distance,gonzalez2021self}.

In order to mitigate the risks associated with error cases, AI models should be accompanied by a catalog of methods that estimate the confidence of the system for each prediction.

\subsection*{Active involvement of radiologists}

Despite Geoffrey Hinton's assessment back in 2016 that deep learning would replace radiologists \citep{alvarado2022should}, it has become clear by now that it will not. Deep neural networks \emph{do} have the potential to reduce the total workload of healthcare professionals and allow for more effective utilization of hospital resources. However, this depends to a great extent on the involvement of human users with the system. This involvement is even more crucial for continual learning algorithms.

First, data needs to be \textbf{curated and annotated} following the format established by the structured template. Part of it can be used for training, fine-tuning the model to the in-house distribution. Even more crucial is maintaining an internal \emph{test} dataset for benchmarking the performance of the system. Labeling does not necessarily entail an additional task as long as the structured template was designed taking into account clinically relevant aspects. In fact, many pieces of SaMD already allow the radiologist to correct the predictions in their interface. Ideally, radiologists onsite should act as a \textbf{quality assurance mechanism} for the AI system. When they note that clear mistakes are being made, the use of the system should be put on hold until these issues are resolved by further training or tuning. The \textbf{utility of the system} should also be assessed by considering additional overhead vs. the benefits.

Of course, maintaining an in-house dataset for benchmarking and manual quality monitoring are tasks that should be carried out for locked just as for lifelong products. In addition, \emph{active learning} methods can produce high-quality annotations with minimal user input \citep{gotkowski2022i3deep}.

\section{Conclusions and Outlook}

Until recently, manufacturers of AI medical software were reluctant to develop products that continue learning post-deployment due to the regulatory overhead involved with obtaining re-certification. Instead, they would stand on large and heterogeneous datasets hoping that models would stay robust for as long as possible. But the risk of progressive performance deterioration  of locked AI cannot be overlooked.

There are valid concerns on whether the current process of AI software approval addresses these risks. \citet{wu2021medical} investigated 130 certified devices and found that 126 only went through retrospective studies, and 93 did not report any multi-center validation. From 54 high-disk devices, none showed prospective evidence. Similarly, \citet{van2021artificial} reported that for 100 commercially available AI products, 192 from 237 studies were retrospective, and only 71 released multi-site results.

\begin{figure*}[h]
\centering
\includegraphics[width=\linewidth]{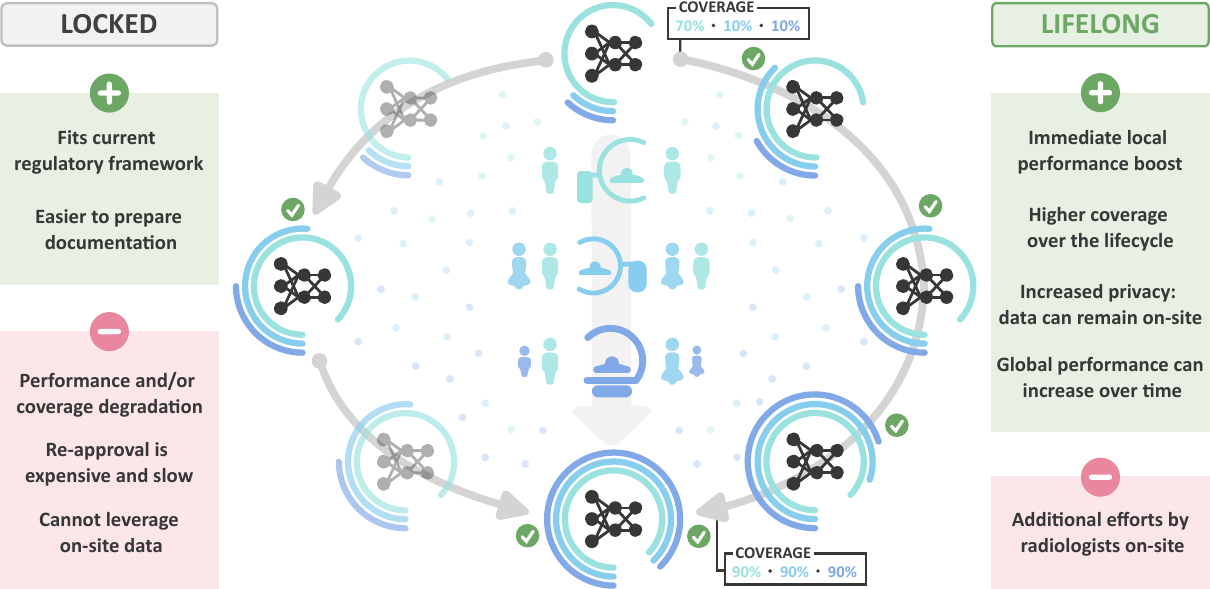}
\caption{Advantages and disadvantages of locked vs. continual learning systems with respect to resource utilization and model performance.} \label{fig:REG_static_vs_continual}
\end{figure*}

Fortunately, we are quickly moving towards a lifecycle regulatory protocol that allows for post-market modifications. Both the European Union through the AI Act and the FDA with the PCCP introduce a similar concept: Planned changes must be described in detail during the approval process and be accompanied by a series of mechanisms that ensure the safety and effectiveness of models through real-world performance monitoring, protection of patient privacy, mitigation of biases, transparency, and traceability. Besides the cases of the USA and EU, other states -- including Canada and the UK -- are drafting regulations that may soon describe how best to handle continual ML products \citep{canadian2022overview}.

Technical advances in continual and active learning, interpretability, and uncertainty estimation are key for maximizing trust and minimizing the additional work that collaborating with AI systems signifies for radiologists. The potential benefits of the new framework include higher diagnostic accuracy, minimization of risks thanks to continuous monitoring, and the possibility to establish a sustainable business model where hospitals do not need to acquire a new product every few years.

While recent guidelines for continuous AI-based diagnostic algorithms provide a solid foundation, they fail to address critical gaps such as the definition of appropriate evaluation metrics, datasets, and performance deterioration thresholds, as well as the handling of performance trade-offs across sub-populations. These challenges align with other limitations identified in the regulation of AI-based medical devices, such as the lack of clear definitions for diagnostic tasks, mechanisms for comparing similar algorithms, and robust characterization of safety and performance, underscoring the need for structured evaluation phases and third-party assessments \citep{larson2021regulatory}. For now, these questions are left to the discretion of the manufacturers and regulatory bodies, with the hope that objective standards are established in the coming years, as the industry matures.

\section{Acknowledgements}

This work was supported by the German Ministry for Health (BMG) with grant [ZMVI1-2520DAT03A].



\bibliographystyle{elsarticle-harv} 
\bibliography{references}

\end{document}